\documentclass[aip,prb,amsmath,amssymb,reprint]{revtex4}

\usepackage{graphicx}% Include figure files
\usepackage{dcolumn}% Align table columns on decimal point
\usepackage{color}
\usepackage{ulem}
\usepackage{mathptmx, helvet, courier}

\newcommand{\rout}{\bgroup\markoverwith%
{\textcolor{red}{\rule[.5ex]{2pt}{0.5pt}}}\ULon}

\begin{document}

\title{Topological transformations in proteins: effects of heating and 
proximity of an interface}

\author{Yani Zhao}
\affiliation{Institute of Physics, Polish Academy of Sciences,
 Al. Lotnik{\'o}w 32/46, 02-668 Warsaw, \\ Poland}

\author{Mateusz Chwastyk}
\affiliation{Institute of Physics, Polish Academy of Sciences,
 Al. Lotnik{\'o}w 32/46, 02-668 Warsaw, \\ Poland}

\author{Marek Cieplak}
\email{mc@ifpan.edu.pl}
\affiliation{Institute of Physics, Polish Academy of Sciences,
 Al. Lotnik{\'o}w 32/46, 02-668 Warsaw, \\ Poland}

\date{\today}

\begin{abstract}
\noindent
Using a structure-based coarse-grained model of proteins,
we study the mechanism of unfolding of knotted proteins
through heating. We find that the dominant
mechanisms of unfolding depend on the temperature applied
and are generally distinct from those identified for
folding at its optimal temperature.
In particular,
for shallowly knotted proteins,
folding usually involves formation of two loops whereas unfolding
through high-temperature heating is dominated by untying of single loops.
Untying the knots is found to
generally precede unfolding unless the protein is deeply knotted
and the heating temperature exceeds a threshold value.
We then use a phenomenological
model of the air-water interface to show that such an interface
can untie shallow knots, but it can also make  knots 
in proteins that are natively unknotted.
\end{abstract}

%\pacs{87.10.+e; 87.15.−v}


\maketitle

\section*{Introduction}

%A non-trivial mathematical knot is a state of a closed curve. 
A closed curve may form a well-defined mathematical knot whose
main characteristic is the number of intersections when projected
onto a plane. Unknotting it would require slicing through it.
A circular DNA forms a closed curve that typically contains knots.
There have been many
studies of knots in such DNA \cite{Buck,DNA,Arsuaga,Stasiak}.
DNA, however, may exist
both in closed and open forms, but in either case all topological
transformations must occur through the action of cutting and reattaching
enzymes such as topoisomerases \cite{Stasiak1,Berger,Flamini} and 
resolvases \cite{Resolvase,Sumners}. The cutting has been observed to be
facilitated by supercoiling that tightens DNA knots \cite{Dietler}.
Topology changes of open DNA also require cutting because of
the large size of the molecule.
%The DNA in an open form is generally
%just too large a molecule for the ends to matter.

Knotted proteins 
\cite{Taylor,Jackson2,Jackson1,Virnau,Virnau1,Stasiakp,Mallam,KnotProt}, 
on the other hand, are small compared to DNA and their
topological states may evolve in time through large conformational
changes such as folding from an unknotted extended state and unfolding.
All of the native protein knots can be obtained by repeatedly twisting
a closed loop and then threading one of the ends through the loop.
Therefore, they are called twist knots.

Theoretical studies have established that the folding behavior
depends on whether the native state of the protein is knotted
in a deep or shallow fashion: it is much harder to tie  the former
than the latter \cite{Wallin,dodging,topTakada,Micheletti,Noel,deep,shallow} 
but the process, in both cases, is predicted to be helped by the
nascent conditions provided by the ribosomes \cite{deep,shallow}.
A protein is considered to be deeply knotted in its native 
state if both ends of the knot, as determined, {\it e.g.} by the KMT 
algorithm \cite{Taylor,km1},  are far away from the termini
(in practice, by more than about 10 residues).
Otherwise it is considered to be knotted shallowly. Notice that the
sequential heterogeneity of a protein positions the knot
in a specific sequential region and tightening of the knot, upon
protein stretching from its termini, goes through jumps to  
specific locations \cite{Sulkowska_2008}.   %,Israel}.

In this paper, we consider two different types of conformational
changes: thermal unfolding and protein deformation induced by
a nearby air-water interface. We find that a sufficiently high
temperature can untie any type of knots if one waits long enough, 
but the topological pathways
of unfolding are generally not the reverse of those found for folding.
The air-water interface may induce unknotting of shallow knots
but we have also found an example of a situation in which a protein
acquires a knot.
We perform our simulations within a structured-based coarse-grained
model and the interface is introduced empirically through coupling
of a directional field to the hydropathy index of an amino acid residue
in the protein \cite{airwater}. Such a field favors the hydrophilic
residues to stay in bulk water and the hydrophobic residues to seek
the air, leading to surface-induced deformation and sometimes
even to denaturation, defined by the loss of the biological functionality.
The simplified character of the model leads to results that are necessarily 
qualitative in nature – they just illustrate what kinds of effects the 
presence of the interface may bring in, especially in the context of the 
topological transformations.

It should be noted that the behavior of proteins and protein layers at the
air-water interface is of interest in physiology and food science.
For instance, the high affinity of lung surfactant proteins to stay 
at the surface of pulmonary fluid generates defence mechanisms 
against inhaled pathogens \cite{Head}. 
The layers of the interface-adsorbed proteins typically show
viscoelastic properties \cite{Graham,Reich,Murray} and the
enhanced surface viscosity of the pulmonary fluid is thought 
to provide stabilization of alveoli against collapse \cite{lungs}.
Protein films in saliva increase its retention and facilitate 
its functioning on surfaces of oral mucosa \cite{saliva}.
Various proteins derived from malted barley have been found to
play a role in the formation and stability of foam in beer \cite{beer}.
Adsorption at liquid interfaces has been demonstrated to lead to
bending of and ring formation in amyloid fibers \cite{Mezzenga}.

There are many theoretical questions that pertain to the 
behavior of proteins at the air-water interfaces. The
one that we explore here is whether the interfaces can affect 
topology. We find that indeed it can: the shallowly
knotted proteins may untie and some unknotted proteins
may acquire a shallow knot. Deeply knotted proteins get distorted
but their knottedness remains unchanged.
We consider four proteins: 1) the deeply knotted YibK 
from {\it Haemophilus influenzae} with the PDB \cite{pdb} structure code
1J85 \cite{1J85}, 2) the shallowly knotted MJ0366 from  methanogenic
archea {\it Methanocaldococcus jannaschi} (PDB:2EFV) -- this is
the smallest knotted protein known, 3) the shallowly knotted DndE 
from {\it Escherichia coli} (PDB:4LRV), and 4)  
chain A of the pentameric ligand-gated ion channel 
from  {\it Gleobacter violaceus} (PDB:3EAM) \cite{Bocquet}
which is an unknotted protein. From now on, we shall refer
to these proteins by their PDB codes. In order to elucidate
the effects of hydrophobicity we shall also consider certain "mutated"
sequences in which certain residues are replaced by other
residues without affecting the native structure.
The proteins 1J85, 2EFV, and 4LRV  have the sequential lengths
of 156, 82 and 107 respectively and the corresponding sequential
locations of their knots are 78-119, 11-73, and 8-99. Thus the knot in 2EFV is
shallow at the C-terminus whereas the one in 4LRV -- at both termini.

\section*{Methods}

Our basic structure-based coarse-grained model of a protein has been 
described in detail in refs. \cite{Hoang2,JPCM,plos,Wolek1}. 
We use our own code. Other implementations of structure-based 
(or Go-like \cite{Go}) models can be found in
 refs. \cite{Takada,models,Clementi,Karanicolas,Paci0,Levy}.
The primary ingredient of the model
is the contact map which specifies which residues may
form non-bonding interactions described by a potential well.
There are many types of contact maps, as summarized in ref. \cite{Wolek},
and we take the one denoted by OV here.
This OV map is derived by considering
overlaps between effective spheres assigned to heavy atoms in the
native state. The radii are equal to the van der Waals radii
multiplied by 1.24~\cite{Tsai}. 
The potentials assigned 
to the contacts between residues $i$ and $j$ are given by
\begin{equation}
V(r_{ij})=\epsilon \; \left[\left(\frac{r_{ij}}{\sigma _{ij}}\right)^{12} \;-\;
\left(\frac{r_{ij}}{\sigma _{ij}}\right)^{6}\right] \;\;\;.
\end{equation}
The length parameters $\sigma_{ij}$ are derived pair-by-pair from the native
distances between the residues -- the minimum of the potential must coincide
with the $\alpha$-C--$\alpha$-C distance. Consecutive $\alpha$-C  atoms
are tethered by the harmonic potential $k_r\;(r_{i,i+1} -r^n_{i,i+1})^2$,
where  $k_r$=100 $\epsilon$/{\AA}$^2$ and
$r^n_{i,i+1}$ is the native distance between $i$ and $i+1$. 
The local backbone stiffness favors the native sense of the 
local chirality, but using the self-organized polymer model \cite{Hyeon}
without any backbone stiffness yields similar results \cite{Wolek1}.

The value of the parameter $\epsilon$ has been calibrated to
be of order 110 pN\;{\AA} which was obtained by making comparisons
to the experimental data on stretching \cite{plos}.
We use the overdamped Langevin thermostat and the
characteristic time scale in the simulations, $\tau$, is of order 1 ns.
The equations of motion were solved by the 5th order
predictor-corrector method. Due to overdamping, our code is equivalent
to the Brownian dynamics approach.
A contact is considered to be established if its length is 
within 1.5 $\sigma$ \cite{JPCM}.
The trajectories typically last for up to 1~000~000 $\tau$. 
%The temperature corresponding to optimal folding kinetics
%in unknotted model proteins is close to $T_r$=0.3 $\epsilon$/{\AA}\;\cite{Wolek1}
%and this is the temperature used in the studies involving the interface.

Despite its simplicity, the structure-based model used here has been
shown to work well in various physical situations. In particular,
it is consistent (within 25\% error bars) with the experimental
results on stretching for 38 proteins \cite{JPCM,plos,models}.
It also has good predictive
powers. For instance, our simulations \cite{JPCM}
have yielded large mechanostability of two cellulosome-related
cohesin proteins c7A (PDB:1AOH) and c1C (PDB:1G1K) that
got confirmed experimentally \cite{Valbuena}. In the case of c7A,
the calculated value of the characteristic unravelling force
is 470 pN and measured -- 480 pN.
The model also reproduces the intricate multi-peak force profile
corresponding to pullling bacteriorhodopsin out of a membrane \cite{Janovjak}.
The equilibrium positional RMSF patterns have been found to be
agree with all-atom simulations, for instance, for
topoisomerase I \cite{Szklarczyk} and Man5B complexed with a
hexaose \cite{Poma}. This model has also been used to study
nanoindentation of 35 virus capsids \cite{virus,virus1}
and to demonstrate that characteristic collapse forces
and the initial elastic constants are consistent
with the experimental data \cite{virology}.

The air-water interface is centered at $z$=0 and extends in the
$x-y$ plane so that the bulk water corresponds to negative
$z$ and air to positive $z$. However, it should be noted that the
interface is diffuse -- its width is denoted by $W$.
The interface-related force acting on the $i$th ${\alpha}$-C
atom is given by \cite{airwater}
\begin{equation}
F_i^{wa}=q_i\;A \;\frac{exp(-z_i^2/2W^2)}{\sqrt{2\pi}W}
\label{hyforce}
\end{equation}
where $q_i$ is the hydropathy index,
$A$ is set equal to 10$\;\epsilon$, and $W$=5 {\AA}.
We use the values of $q_i$ as determined by Kyte and Doolittle \cite{Kyte}.
They range between --4.5 for the polar and charged ARG and 4.5 for the
hydrophobic ILE.  Other possible scales are listed in ref. \cite{Palliser}.
For each protein, we can identify a degree of hydrophobicity $H$ in terms the values of $q_{i}$ of its amino acids,
$H\;=\;\frac{1}{N}\;\sum_{i=1}^N \; q_{i} $,
where the sum is over the amino acids in the protein.
Properties of protein conformations are assessed by the
fraction, $Q$, of the native contacts that are present in the conformation.

The phenomenologically motivated addition of the air-water
term to the basic structure-based model
leads to the experimentally observed formation of a protein
layer \cite{Leheny} at the interface and gives rise to the in-layer
diffusive behavior which is characteristic of soft colloidal glass
with the intermediate values of the fragility indices \cite{airwater}.
Specifically, as a function of the number density of the proteins at the
interface, the surface diffusion coefficient obeys a Vogel-Fulcher-Tamann
law. This is consistent with the microrheology experiments on
the viscoelastic behavior of protein layers \cite{Cicuta2003,Leheny2}.

In the initial state, $N_p$ proteins ($N_p$ is between 2 and 50 are
placed in a large square box so that their center
of mass are around  $z$=--3.2 $\pm$ 0.4 nm
with the $x$ and $y$ coordinates selected randomly.
Their initial conformations are native.
The box is bounded by a repulsive bottom at -7 nm and by repulsive sides.
The force of the wall-related repulsion decays as the
normal distance from the wall to the tenth power.
The walls may be brought to a desired smaller
separation in an adiabatic way, however, here we focus
on the dilute limit in which the proteins are far apart.
The purpose of considering many proteins
simultaneously is to generate statistics of behavior and a spread
in the arrival times to the interface.
If the proteins happen to come close to one another, their
mutual interactions correspond to repulsion that forbids overlap.

The thermodynamic properties of the system in the bulk are assessed by 
determining the temperature ($T$) dependence of $P_0$ -- the probability
of all contacts being established simultaneously. $P_0$ is determined in
several long equilibrium runs. 
For typical unknotted proteins, the optimum in the folding time
is in the vicinity of $T=T_r=0.3\;\epsilon /k_B$ ($k_B$ is the Boltzmann
constant) where $P_0$ is nonzero. A more detailed discussion
of this point is presented in ref. \cite{Wolek1}. $T_r$ then
plays the role of the effective room temperature. This value of $T$ is also
consistent with the calibration of the parameter $\epsilon$.

Thermal unfolding is studied by considering a number of trajectories 
at $T > T_r$ that start in 
the native state and last for up to  1~000~000 $\tau$. Unfolding is
achieved if all native contacts that are sequentially separated 
by more than a threshold value of $l$
residues are ruptured for the first time \cite{deep}.  
An ideal unfolding would involve breaking of all contacts, but
such simulations would take unrealistically long to run. We thus
introduce the threshold that separates
contacts that are sequentially local from the non-local ones.
Contact in $\alpha$-helices do not exceed the distance of 4.
Usually, we take $l$=10.
The median value of this rupture time defines the characteristic
and $l$-dependent unfolding time $t_{unf}$.
An alternative criterion could involve crossing a threshold
value of $Q$.

The dynamics of staying in the
knotted state in the bulk or on approaching the air-water interface
is assessed by monitoring the time dependence
of $P_k(t)$ -- the probability that, at time $t$, the protein stays
in its native topology.

\section*{Results}

\subsection*{Thermal unfolding}

Even though the deeply knotted 1J85  protein is difficult to fold
from a fully extended conformation at any $T$, we find that
it is easy to unfold it at elevated $T$, if the waiting time is
sufficiently long. Within our cutoff-time, we could observe it happen for
$T \ge 0.85 \epsilon /k_B$. For $T \ge 1.0 \epsilon /k_B$ we have not
recorded any refolding events after full unfolding. At $T=0.85 \; \epsilon /k_B$,
21\% of the 28 trajectories resulted in retying the trefoil knot.
Note that the starting conformation for the refolding process is not
at all fully extended and is thus biased towards knotting -- a situation
most likely encountered in ref. \cite{dodging}.
The loss of all contacts may result in conformations
that look like expanded  globules.
Taking $l$ of 10, the values of $t_{unf}$ are 565~045, 116~512, and 25~479 $\tau$
for $T$ equal to 0.9, 1.0, and 1.2 $\epsilon/k_B$ respectively.
Breaking contacts %exceeding a sequential threshold of 10 sites
is not directly related to untying. We find that the median untying 
times are 198~850, 85~050, and 34~710 $\tau$ for the same
temperatures respectively. This indicates that at the lower two of the
three temperatures untying precedes unfolding and decreasing the $l$ enhances the
gap between the two events (see Fig.~S1 in Supplementary Information, SI).
Only in one trajectory out of the total of 25 at
$T=\;1.0\epsilon/k_B$, unfolding takes place 200 $\tau$ earlier than untying.
For $T=1.2\;\epsilon /k_B$, unfolding takes place before untying
in most of the trajectories if one takes $l$=10, but for $l$=4, the
reverse holds. It is only at $T=1.5\;\epsilon /k_B$ that unfolding
always takes place before untying even if $l$=4.
%For instance at $T=\;1.2\epsilon/k_B$, unfolding occurs earlier in $64\%$ of the trajectories for $l=10$, and their $t_{unf}$ is 567 $\tau$ larger than the untying time averagely. 

The unfolding pathway of knotted proteins has been studied
in ref. \cite{unfoldsulko}
for a structurally
homologous YibK-like methyltransferase (PDB:106D). The theoretical
part of the study also involved 
a structure-based model, but with a very different contact map.
The finding was that untying takes place after unfolding and this
was taken as a signature of a certain hysteresis in the process.
However, the value of $T$  was not specified -- presumably
the simulations were done at a high $T$. We just demonstrate that
the actual sequence of the unfolding events depends on $T$. Since,
in our model, $T$ of $1.0\;\epsilon /k_B$ corresponds to about 850 K,
it is the still lower $T$ that are relevant experimentally
and thus observing unfolding before unknotting on heating seems unlikely.
However, the experimental studies 
%(see also a related analysis of the MTase protein \cite{Jennings}) 
involve chemical denaturation
by Gnd-HCl, which allows for a broader range of conditions that
are meaningful experimentally.

\begin{figure}[h]
\centering
\includegraphics[width=0.45\textwidth]{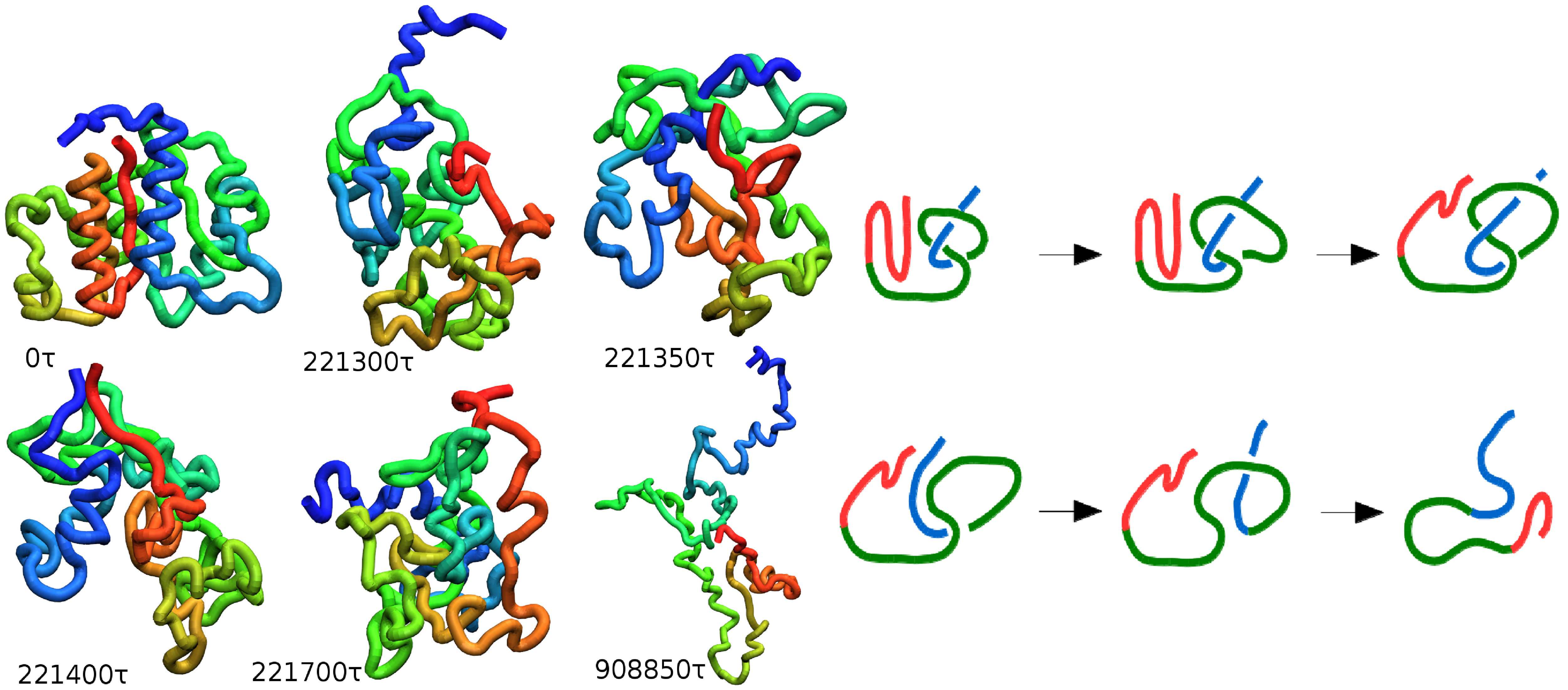}
\caption{An example of thermally induced unfolding of 1J85 through the DT 
mechanism at $0.85 \; \epsilon /k_B$.
The six panels on the left  show successive snapshots of the backbone
conformations at times indicated. The six panels on the right
provide the corresponding schematic representations of these conformations.
The N-terminal segment is shown in shades
of orange and red, the C-terminal segment in shades of blue, 
and the middle segment in shades of green.
} \label{tunf1j85dt}
\end{figure}

\begin{figure}[h]
\centering
\%%includegraphics[width=0.48\textwidth]{Fig2b}
\includegraphics[width=0.45\textwidth]{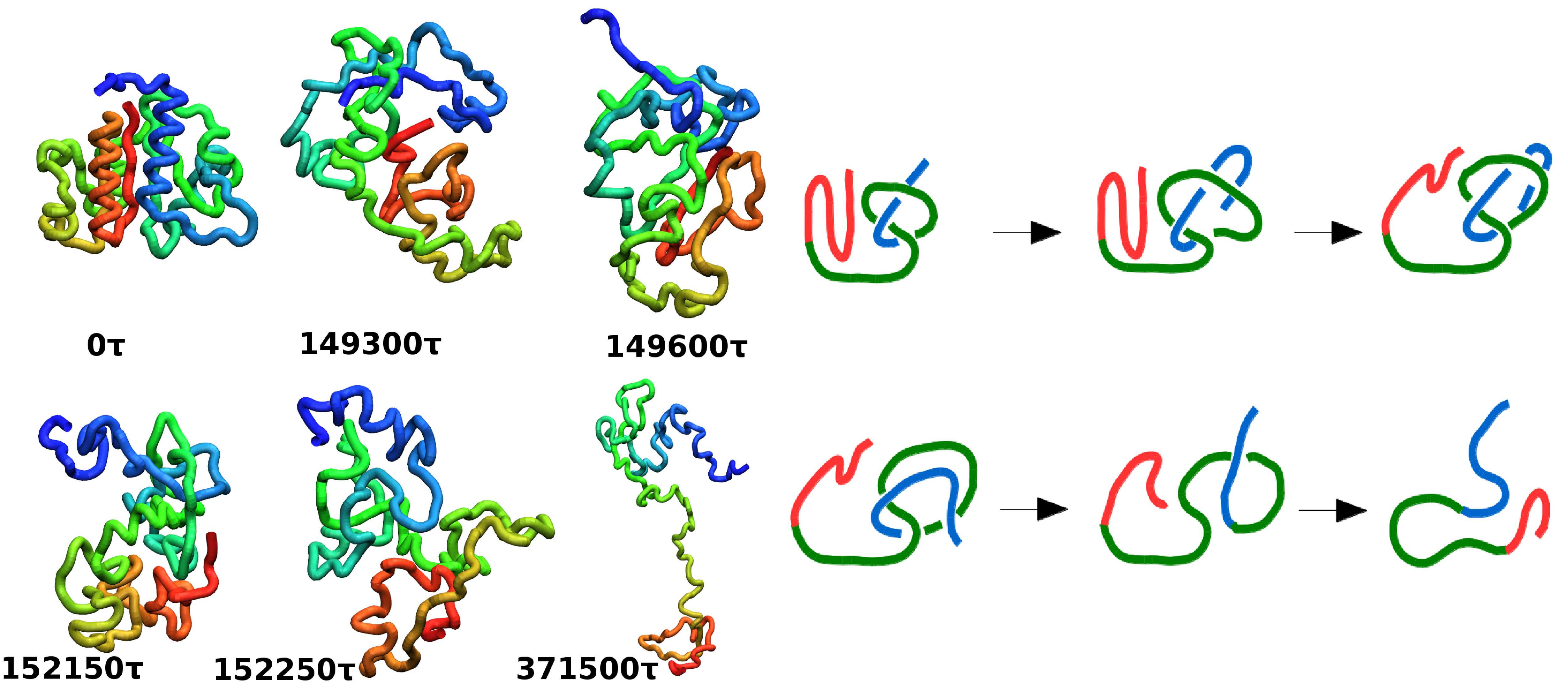}
\caption{Similar to Fig. \ref{tunf1j85dt}, but unfolding proceeds through
the SK mechanism and the trajectory corresponds to $T=0.9 \; \epsilon/k_B$.
} \label{tunf1j85sk}
\end{figure}

The mechanisms of unknotting in 1J85 are dominated by direct 
threading (DT) events, 
illustrated in Fig. \ref{tunf1j85dt}, followed 
in statistics by slipknotting (SK) events, 
as illustrated in Fig. \ref{tunf1j85sk}. 
We observe no other unfolding mechanisms.
They have been discussed in refs. \cite{dodging,shallow}
in the context of folding except that now they operate in reverse. For instance,
the DT mechanism involves pulling of a terminus of the protein out of a loop and the SK
mechanism involves pulling a slipknot out of the loop. The determination
of the precise nature of the process is based on a visual monitoring
of the subsequent snapshots of the evolution.
The exact proportions between the mechanisms depend
on the $T$.
The red color is used for the N-terminal segment, blue -- for the
C-terminal one and green -- for the middle part of the backbone.
However, the number of instances of unknotting 
through SK decreases with a growing $T$ (32\%, 8\%, and 0\%  at 0.9,
1.0 and 1.2 $\epsilon /k_B$).
Unknotting in the trajectory shown in Fig. \ref{tunf1j85dt}
takes place at time 221~400 $\tau$
so the last panel corresponds to a situation in which the
protein is unknotted but not yet fully unfolded.

Topological pathways of folding in the shallowly knotted  2EFV have been 
demonstrated to be of two basic kinds: through single loops \cite{Micheletti}
or through two smaller loops \cite{shallow}. The latter is the
dominant pattern and is a two-stage process. 
The two-stage pathways have not been observed in the deeply knotted 1J85.
In each of these cases, the
specific mechanisms of making the knot involve, in various proportions,
DT, SK, and mouse-trapping (MT). MT is similar to DT but the knot-loop
moves onto the terminal segment of the protein instead of
the other way around.
There is also a possibility of an embracement (EM) \cite{shallow} in which 
a loop forms around a terminal segment.
The DT, SK, MT, and EM mechanisms may operate either
at the level of a single larger loop in a process, which is
topologically one-stage, or at the level of two
smaller loops and hence in two stages. Again, the identification
of the nature of the pathway is obtained visually.

\begin{figure}[h]
\centering
\includegraphics[width=0.48\textwidth]{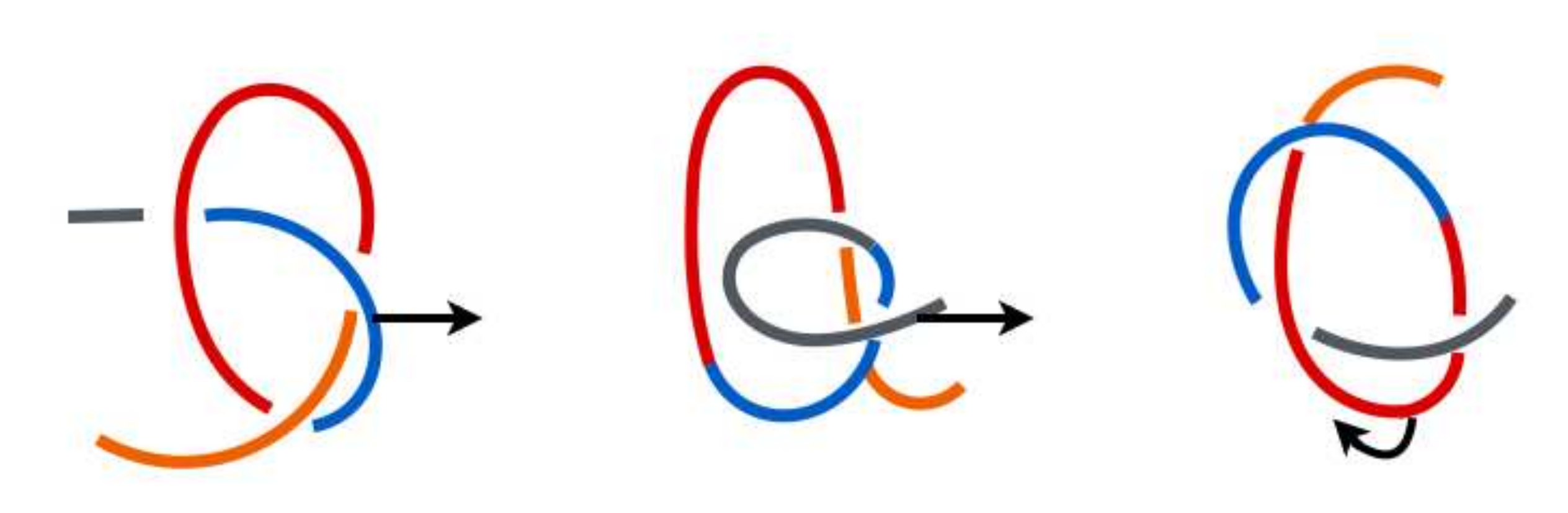}
\caption{A schematic representation of a single-loop DT (left), SK (middle) and MT-based (right) 
thermal unfolding of 2EFV at $T=0.7\; \epsilon/k_B$. 
} \label{tunf2efv}
\end{figure}

When unfolding 2EFV at $T=0.5\; \epsilon /k_B$, all events are two-stage,
exclusively SK-based, and are soon followed by refolding. At 0.7 $\epsilon /k_B$,
only 28\% of 50 trajectories are two-stage 
(DT and SK are involved in each stage) and the 
remaining ones are one-stage. Most of them refold back soon afterwards.
At $T \ge \; 1.0 \epsilon /k_B$ there is no refolding and all trajectories
unfold through the single loop mechanism. The process is dominated
by the DT events, followed by SK, and then some MT ones. An example of a
DT-based pathway is shown in Fig. \ref{tunf2efv}.
The N-terminal segment (1-16) is marked in orange, sites 17-53 in red,
sites 54-78 in blue, and the C-terminal segment (79-82) in gray.
%Untying in this trajectory takes place around 149 300 $\tau$
%and unfolding at 180 605 $\tau$ (the last panel is for an
%unfolded conformation). 
In all trajectories, untying
of 2EFV occurs before thermal unfolding
(for $4\le l\le 10$ at $T \le \; 1.5 \epsilon /k_B$ -- 
see Fig.~S1 in SI). 
%This is similar with what we have observed for the deeply knmotted 1J85.
%The time-dependence of the probability of being in an unfolded state,
%$P_{unf}$ for several values of $l$ is shown in Fig. \ref{punf}
%in SI and compared with the behavior of
%$P_k(t)$.

The physics of folding and unfolding in 4LRV is found to be similar
to that of 2EFV, but the DT unfolding events are more likely to
proceed from the N-terminus instead of the C-terminus. Another difference is 
that folding at $T_r$ is seen to take place exclusively through
the two-loop mechanism. 12 out of 50 trajectories led to folding.
7 of them proceeded through the EM-SK pathway, 4 through SK-SK,
and 1 through DT-SK (see Fig.~S2 in SI).

We conclude that the thermal unfolding processes of the
knotted proteins are generally distinct from a
simple reversal of folding. For instance, the dominant two-loop
folding trajectories do not form a reverse topological template 
for the dominant single-loop 
unfolding trajectories.
A similar observation has been already made for unknotted proteins
although it involves no changes in the topology \cite{thermal}.

\subsection*{Knot-untying by the air-water interface}

\begin{figure}[h]
\centering
\includegraphics[width=0.45\textwidth]{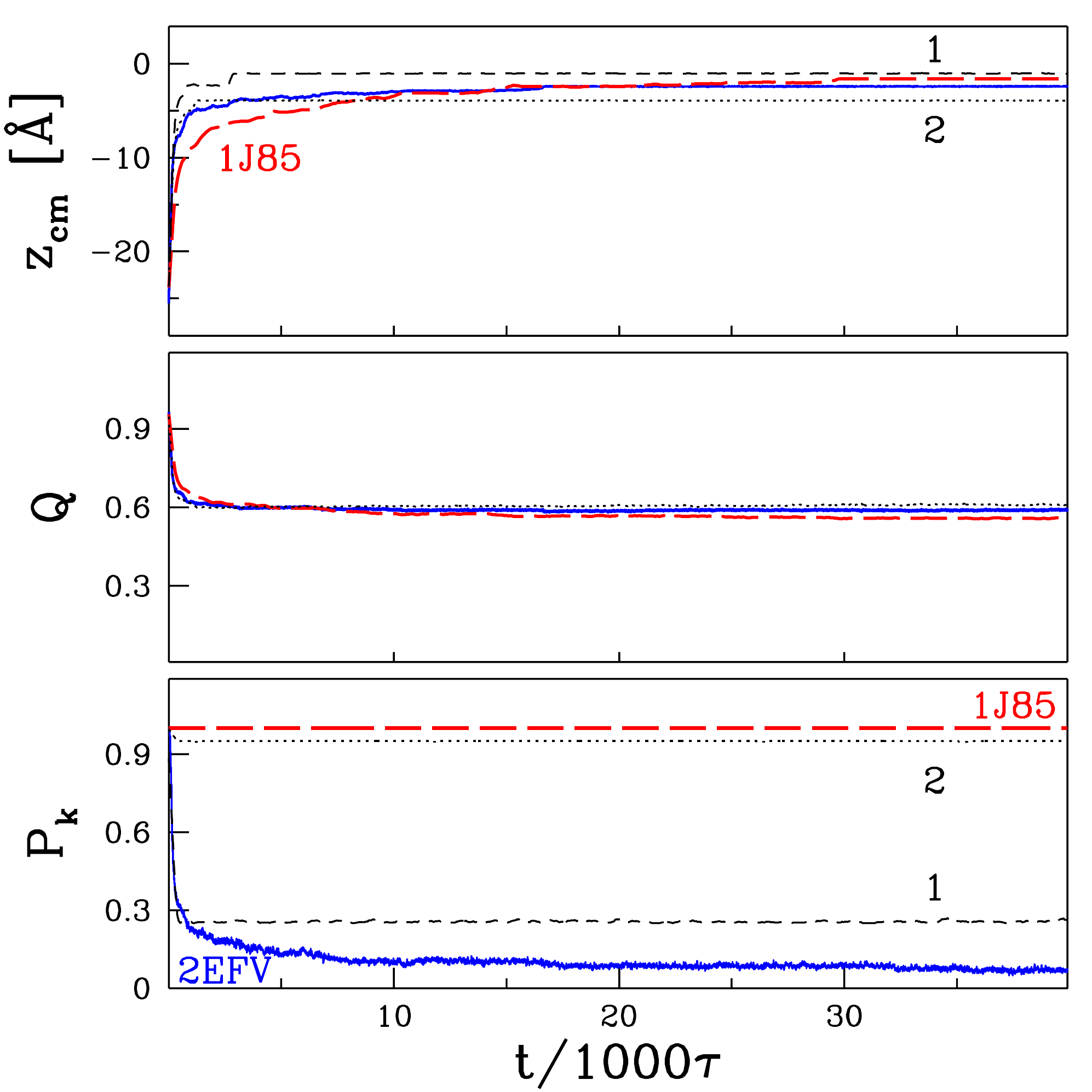}
\caption{ Distance to the surface ($z_{cm}$), 
fraction of preserved native contacts ($Q$) and 
probability of being knotted ($P_k$) as a function of time, $t$, for 1J85 
(red), 2EFV (blue) and its two mutants (1,2; black) at the air-water interface.
The data are based on one trajectory in each case.
} \label{intpro}
\end{figure}

\begin{figure}[h]
\centering
\includegraphics[width=0.450\textwidth]{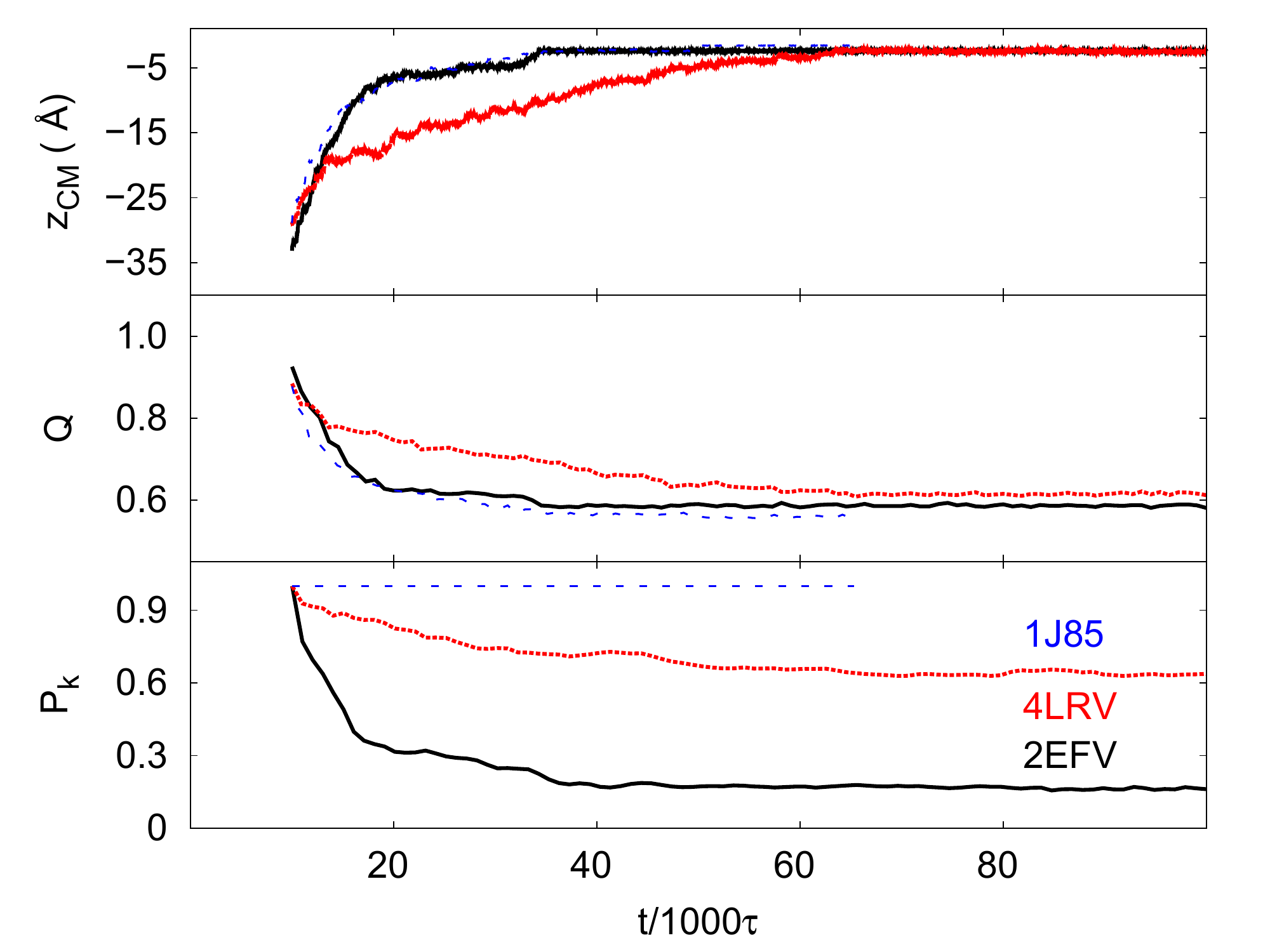}
\caption{Time-evolution of $z_{cm}$, $Q$ and $P_k$ averaged over
50 proteins at $T=T_r$. The black, red, and blue lines are for 2EFV, 
4LRV, and 1J85 respectively.  During the first 10 000 $\tau$,
the proteins diffuse around without the interface. 
} \label{intpro50}
\end{figure}

We now consider the interface-related effects at  $T=T_r$.
The proteins that come to the interface get deformed and lose some
of their native contacts. We find that these phenomena do not
affect the topological state of the deeply knotted 1J85 as demonstrated
in Fig. \ref{intpro}. The data shown are for one example trajectory
which corresponds to a specific starting protein orientation with 
respect to the interface. Various orientations and different initial locations
yield various adsorption times. When one averages over 50 proteins, one gets
the results shown in Fig. \ref{intpro50}. The loss of contacts is
related to the approach of the  center of mass of the
protein(s) to the center of the interface.
The knot-ends may shift from one trajectory to another, but there
is no knot untying. Furthermore combining the effects of the
interface with those of an elevated temperature is found not to 
promote any untying.
%%Besides, no unfolding is observed by taking $l=10$ for trajectories last up to 1 000 000 $\tau$ at $T\le 1.5 \epsilon/k_B$. The adsorption to the interface denaturated the protein and reshape it with a manner of its hydrophobicity, which limits its  movement under thermal fluctuations. For example at  $T=1.5 \epsilon/k_B$, there are averagely 31 native contacts, which are sequentially separated by more than $l=10$ residues, are remained until the end of the simulation.
%Under this condition, stronger fluctuations of the protein at higher
%temperature makes the threading of one terminal out of the loop more
%difficult.

\begin{figure}[h]
\centering
\includegraphics[width=0.5\textwidth]{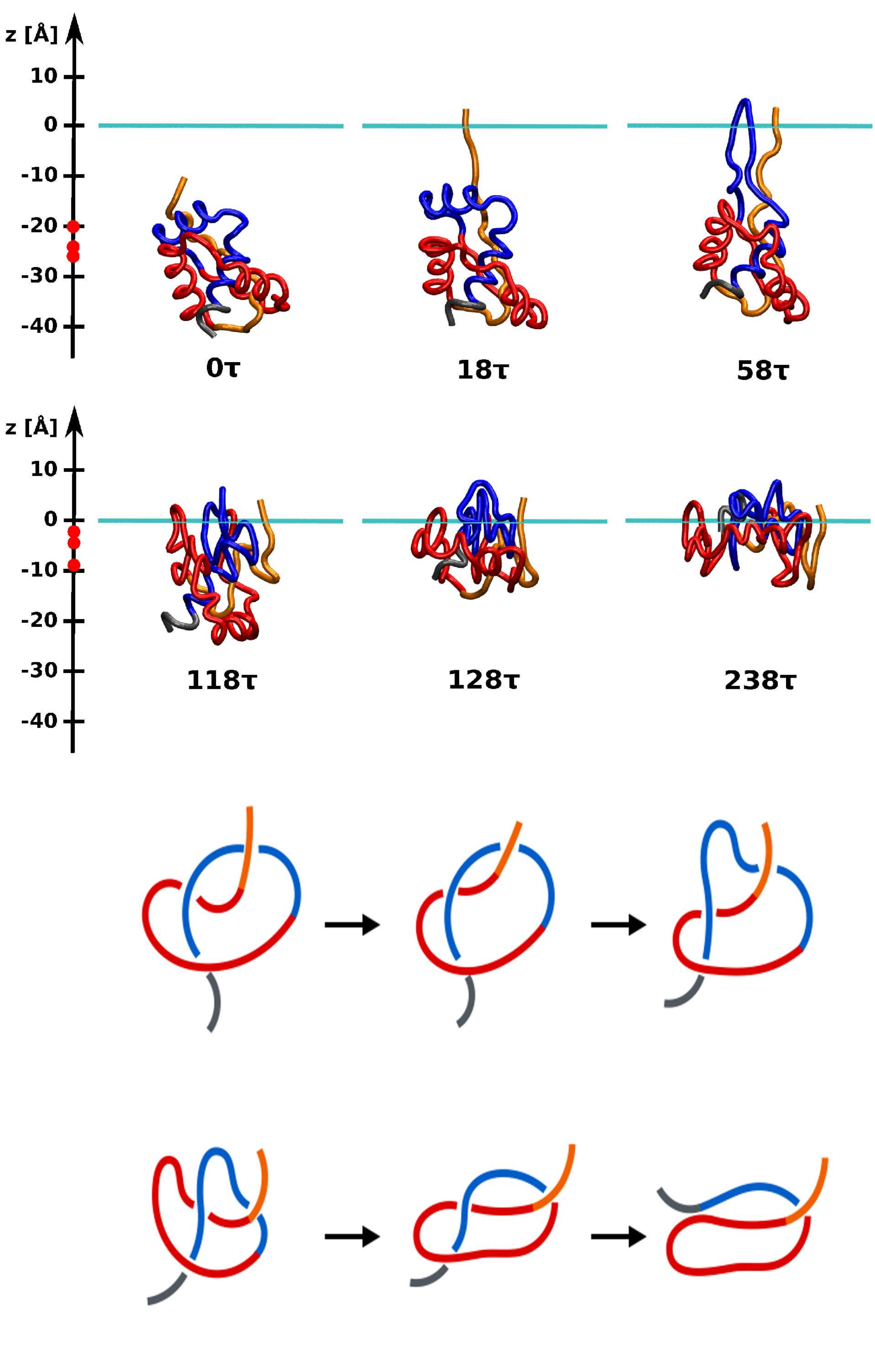}
\caption{ The interface-induced untying of the knot in 2EFV.
The 6 top panels show snapshots of the backbone conformations
at times indicated. The horizontal line shows the level corresponding
to the center of the interface. The first panel corresponds to the 
protein that is still away from the interface. The red marks on the $z$-xis
indicate the subsequent positions of the center of mass of the protein.
The 6 bottom panels show the schematic topological representations 
corresponding to the conformations  shown in the top panels.
Untying is accomplished through the DT mechanism.
%The N- and C-terminus of proteins are colored orange and gray, respectively.
} \label{figwa2efv}
\end{figure}

The situation changes for the shallowly knotted proteins. Now the
knots do untie. An untying process is illustrated
in Fig. \ref{intpro} (2EFV and two of its mutants), 
Fig. \ref{intpro50} (2EFV and 4LRV) and Fig. \ref{figwa2efv} (2EFV). 
Adsorption of 2EFV is driven by the hydrophobic N-terminus 
(its first two residues are hydrophobic while the hydropathy
indices of the first 10 residues add up to $-1.11$) but the
untying process takes place primarily through DT  (7\% by MT)
at the hydrophilic C-terminus.
Due to the distortion of the whole protein, it is difficult 
to decide whether the unfolding
process involves one or two loops so we do not provide the partitioning
numbers.

The last nine residues in 2EFV are \mbox{LNCELVKLD} and their hydropathy 
indices add up to +0.41.
However, the protein can tie back again either
through DT or SK and hence $P_k$ in Fig. \ref{intpro} decays 
to a finite value instead of to zero.
Overall, the changes in the topology, as described by $P_k$, depend
both on the approach to the interface and on the related loss of
the contacts.

We now consider two mutations at the C-terminus in 2EFV. 
The first mutation replaces the last 9-residue
sequential segment by \mbox{LACALVALA} which makes it more hydrophobic
-- the hydropathy indices add up to +2.81 --
and the second mutation, to \mbox{PNPEPPKPD},
makes it hydrophilic -- the hydropathy indices add up to -2.49.
Fig. \ref{intpro} shows that both mutations enhance the probability
of staying knotted but mutation 2 is much more effective in doing so.
The hydrophobic C-terminus of the first mutation favors an accelerated
adsorption with less time to untie. The hydrophilic C-terminus,
on the other hand, gets stuck in the water phase which
preserves the knotted topology of the protein.
In conclusion, the distribution of the hydrophobicity of a knotted protein
is a factor contributing to the untying probabilities at the interface.

A similar behavior is observed for 4LRV (Fig. \ref{intpro50}) except that this
protein is more likely to stay knotted than 2EFV. The two proteins are quite
comparable in their linear size in the native state:
the radius of gyration for 4LRV is 13.08 {\AA}, 
and for 2EFV -- 12.89 {\AA}. However, they differ in the contact-mediated
connectivity significantly: 4LRV has 36\% more contacts than 2EFV. 
This feature makes 4LRV  harder to untie than 2EFV.
Two examples of the interface-induced unknotting of 4LRV are shown
in Fig.~S3 in SI, which demonstrates  
two available untying mechanisms of 4LRV, {\it i. e.} DT and MT with DT occuring
more frequently. SK is not observed in the unknotting of 4LRV, which
may be due to the fact that the terminal outer segments of 4LRV are too 
short to form a slipknot.

\subsection*{Knot-tying at the air-water interface}

\begin{figure}[h]
\centering
\includegraphics[width=0.4\textwidth]{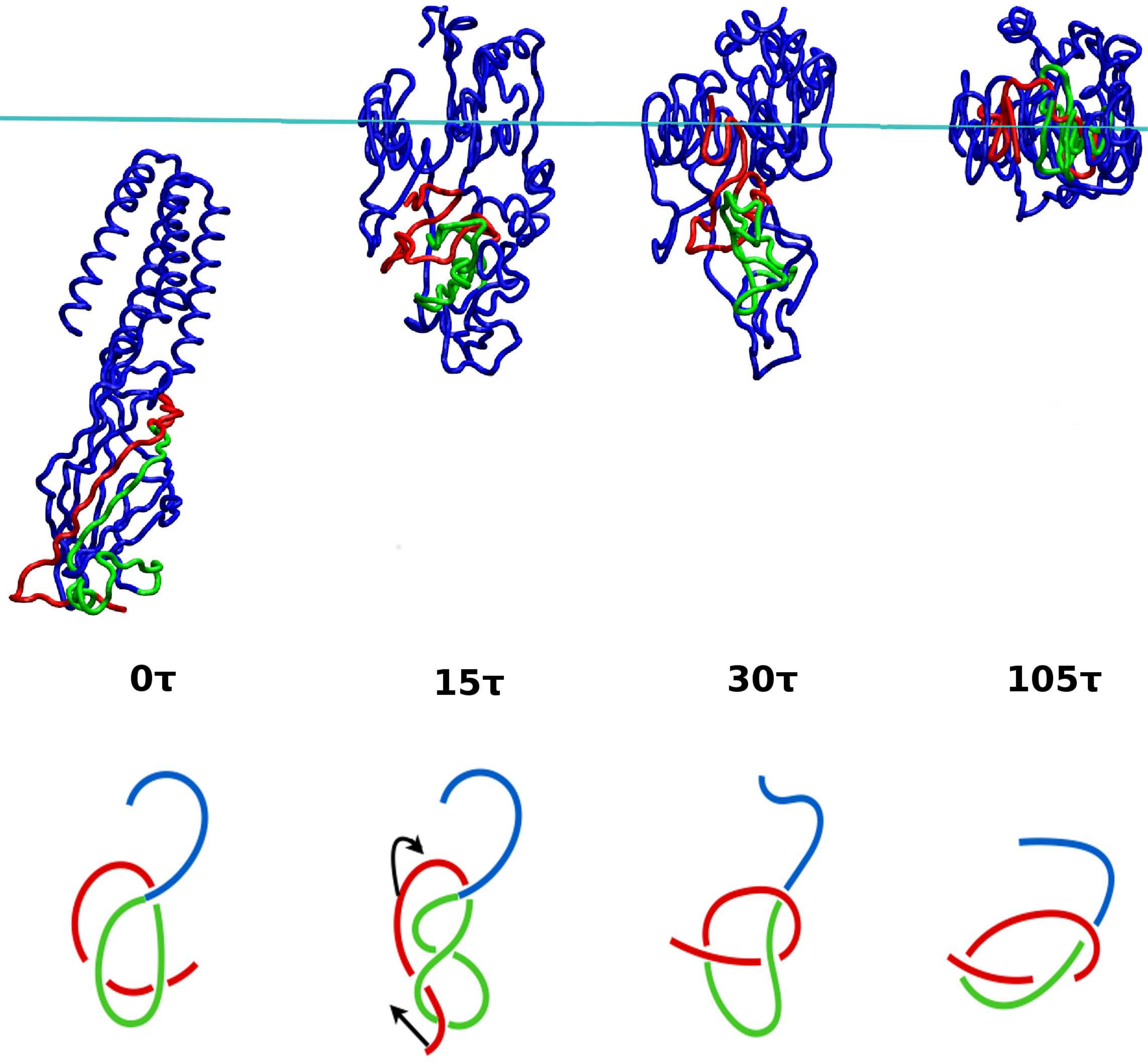}
\caption{An example of the interface-induced knotting process in 3EAM at $T_r$.
The red segment extends from the N-terminus (begins from site 5 since the first four amino acids are not available in the crystal structure) to site 35, 
the green segment -- from 36 to
70, and the blue segment -- from 71 to the C-terminus.
} \label{3eamknot}
\end{figure}

If at least one of the terminal segments 
of an unknotted protein
is hydrophobic, there is 
a possibility that dragging it towards the interface may lead to
formation of a knot. This is, in fact, what we found to
happen in protein 3EAM with $H=+0.32$. This protein comprises 311 structurally
resolved residues.
Its native state is unknotted and thermal fluctuations in the
absence of the interface do not lead to any knot-tying in the
$T$-range between 0.3 and 0.7 $\epsilon /k_B$.
The net hydropathy score for its N-terminal segment of 8  residues, 
which should cross an entangled region of this protein to form a knot, is $+0.41$. 
Due to the low hydrophobicity of this segment, 
the knotting process in most trajectories is accomplished 
when the N-terminus is still  in the water phase, 
as shown in Fig~\ref{3eamknot}. This terminus gets 
lifted to the interface together with other segments 
after the C-terminus (its net hydrophathy of 8 residues is
$+3.35$) of the protein is already adsorbed to the interface.

We find that in 52\% of 50 trajectories, a knot
forms through DT. An example of a formed trefoil knot is illustrated in Fig. \ref{3eamknot}.
If one makes the N-terminal segment more hydrophobic
(to $H$=+2.44) through mutations, then the success in tying the knot
is: 68\%. If one makes it more polar (to
$H=-1.50$) then the success rate is 66\%.  Thus both
mutations increase the knotting probability of 3EAM.
The N-terminus of the hydrophobic mutation can be easily 
adsorbed across the entangled part to the interface, which 
increases the probability of forming a knot. 
On the other hand, if the N-terminal segment is made more
hydrophilic it may be dragged downward to the water phase
after the whole protein gets adsorbed.
This phenomenon may create another chance at passing through the
entangled part of the protein, increasing  the probability of knotting.
The knotted conformation need not last -- the knot, if very shallow, 
may untie through the subsequent evolution.

We have also observed knotting in a transport
protein 4NYK from {\it Gallus gallus}. However, we have not detect it
in other plausible candidates such as 1A91, 1FJK, 1H54, 1H7D, 1KF6, 1N00, 1NNW, 
1O4T, 1RW1, 1YEW, 2EC8, 2FV7, 2HI7, 2I0X, 2IVY, 2OIB, 3JUD, 3KLY and 3KZN.
These proteins have been selected so that at least one of their
termini is hydrophobic since such terminal segments have an
enhanced probability of moving through the protein on
approaching the interface.

%We have surveyed {\color{red} \sout{about}} 30 unknotted proteins and 
%{\color{red} observed knotting \sout{identified a similar knotting behavior}
%also} in a transport 
%protein 4NYK from {\it Gallus gallus}. 
%%The hydrophobicity of this protein is $H=-0.23$.
%The common feature of those two proteins is that 
%they are relatively large which facilitates formation
%of a knotting loop through entanglement. 
%Another is that one of their termini has 
%a high hydropathy score which allows it crosses the entangled 
%zone of the protein. 

\section*{Conclusions}

We have demonstrated that the forces associated with the air-water 
interface may affect the topological state of a protein.
It is an interesting question to ask how to devise experimental
ways to detect such transformations, if they indeed arise.
After all, our model is coarse-grained and phenomenological,
especially in its account of the interface. Thus, further investigation 
such as the comparison between atomistic and
coarse-grained models would be required.
All atom simulations of air-water interfaces,
even in the absence of any proteins, are expected be
complicated due to the huge number of molecules needed to set up
a necessary density profile that would be stationary.
Our simple model  points to possible topological transformations
that may take place at the interface. We hope it will provide
motivation for studies by other means.

It should be noted that topological transformations can also
occur in the intrinsically disordered proteins simply as a result of
time evolution. This has been demonstrated through all-atom
simulations for polyglutamine chains of a sufficient length \cite{Gomez}.
For 60-residue chains, about 10\% of the statistically
independent conformations have been found to be knotted.
These knots can be shallow or deep and are not necessarily
trefoil. The knotted character of these conformation may be related
to the toxicity of proteins involved in Huntington disease  \cite{proteasome}.

Contrary to the results reported in ref. \cite{unfoldsulko}, we find that
shallow knots always untie before the unfolding on heating
and the untying of deep knots may follow unfolding
only at unrealistically high temperatures, though perhaps at acceptable
concentration of the denaturant.
It should be noted that homopolymers without any attractive contact 
interactions may tie knots purely entropically.

\vspace{0.5cm}

\noindent {\bf Acknowledgments}

\noindent
We appreciate comments of \`{A}. G\'omez-Sicilia about the manuscript.
The project has been supported by
the National Science Centre,
Grant No. 2014/15/B/ST3/01905 and by the EU Joint Programme in Neurodegenerative
Diseases project (JPND CD FP-688-059) through the National Science Centre 
(2014/15/Z/NZ1/00037) in Poland.
%umowa podpisana 21 lipca 2014

\section*{Author Contributions}

\noindent
All authors wrote the paper. Marek Cieplak designed the research and wrote the
code. Mateusz Chwastyk wrote knot-related pieces of the code and made
preliminary runs. Yani Zhao did the calculations and made the figures.

\section*{Additional Information}

\noindent
{\bf Competing financial interests}

\noindent
The authors declare no competing financial interests.

\vspace*{1cm}

\end{document}